\documentclass{article}

\usepackage{PRIMEarxiv}

\usepackage[utf8]{inputenc} 
\usepackage[T1]{fontenc}    
\usepackage{hyperref}       
\usepackage{url}            
\usepackage{booktabs}       
\usepackage{amsfonts}       
\usepackage{nicefrac}       
\usepackage{microtype}      
\usepackage{lipsum}
\usepackage{fancyhdr}       
\usepackage{graphicx}       
\usepackage[table,xcdraw]{xcolor}
\usepackage{amsmath}
\usepackage{graphicx} 
\usepackage{booktabs} 

\title{Fairness in LLM-Generated Surveys

}

\author{
  Andrés Abeliuk\thanks{Corresponding author:\textit{ aabeliuk@dcc.uchile.cl}},
  Vanessa Gaete \\
  Department of Computer Science, University of Chile\\ 
  National Center for Artificial Intelligence (CENIA) \\
  Santiago, Chile\\
   \And
  Naim Bro \\
  School of Government, Adolfo Ibáñez University\\
  Millennium Institute for Foundational Research on Data (IMFD) \\
  Santiago, Chile\\
}

\begin{document}
\maketitle

\begin{abstract}
Large Language Models (LLMs) excel in text generation and understanding, especially in simulating socio-political and economic patterns, serving as an alternative to traditional surveys.  However, their global applicability remains questionable due to unexplored biases across socio-demographic and geographic contexts.
This study examines how LLMs perform across diverse populations by analyzing public surveys from Chile and the United States, focusing on predictive accuracy and fairness metrics. The results show performance disparities, with LLM consistently outperforming on U.S. datasets. This bias originates from the U.S.-centric training data, remaining evident after accounting for socio-demographic differences.
In the U.S., political identity and race significantly influence prediction accuracy, while in Chile, gender, education, and religious affiliation play more pronounced roles. 
Our study presents a novel framework for measuring socio-demographic biases in LLMs, offering a path toward ensuring fairer and more equitable model performance across diverse socio-cultural contexts.
 %
\end{abstract}

\section{Introduction}

Large Language Models (LLMs) have launched a disruptive era in natural language processing, with applications ranging from conversational agents to predictive analytics. Models such as OpenAI's GPT series have demonstrated remarkable text generation and understanding capabilities, excelling at tasks beyond their initial training goals~\cite{chatGPT}. These breakthroughs have catalyzed research into the boundaries of LLMs, envisioning diverse applications in fields such as education, healthcare, and social sciences~\cite{exploracion,bailcsssia,ziems2024can}.

A particularly compelling application of LLMs is their ability to simulate socio-political and economic patterns, providing an alternative to traditional surveys and predictive models. Research has demonstrated that LLMs can generate human-like responses based on socio-demographic characteristics, enabling the creation of synthetic datasets whose distributions closely mirror those of human responses to real survey questions~\cite{argyle2023out,kim2024aiaugmented}. These capabilities have been used to predict voting preferences, assess public opinion, and analyze political group perceptions~\cite{lippert2024can}, as well as to replicate individual and collective behaviors to study social dynamics~\cite{park2024generative}. LLMs have been further explored as simulated economic agents, effectively replicating key findings from behavioral economics and social science experiments with an accuracy comparable to human forecasters~\cite{horton2023large,dillion2023can,aher2023using,ashokkumar2024predicting}.


Despite these advancements, limitations and challenges persist. While studies have highlighted the ability of LLMs to serve as proxies for specific human subgroups~\cite{argyle2023out}, they also reveal disparities in model performance across populations, ethical concerns, and the risks of over-reliance on synthetic data~\cite{wang2024large, santurkar2023whose}. Furthermore, most LLM research remains concentrated on the U.S. context, raising critical questions about the generalizability of these models~\cite{myung2024blend,navigli2023biases}. The opaque nature of LLM training data, which spans languages, regions, and socio-demographic groups, makes it difficult to determine whether their exceptional performance in well-represented areas such as the U.S. can be replicated in underrepresented countries or groups.

Ensuring that machine learning models are fair and unbiased is essential for their ethical application in social science research. Most studies that utilize LLM agents simulate human-like responses by choosing from a limited set of predefined options, effectively framing these tasks as classification problems~\cite{kim2024aiaugmented}. Therefore, employing established fairness metrics from machine learning to analyze outcomes offers a systematic approach to identifying and addressing inequalities in the context of LLMs~\cite{verma2018fairness,barocas2023fairness}. 

These metrics typically focus on group fairness, which ensures equitable treatment across different socio-demographic groups~\cite{mehrabi2021survey,caton2024fairness}. Group fairness can be further divided into between-group fairness, which evaluates differences between distinct subgroups. By applying these metrics, it becomes possible to assess how LLMs reflect soci-demographic groups, ensuring that the models do not perpetuate existing societal biases or inadequately represent specific populations.

Building on these considerations, this study measures Fairness and adequate representation of diverse socio-demographic and geographic groups in  LLMs, with a specific focus on Chile. Chile serves as a compelling case study for examining inequalities in LLM representation and performance across geographic contexts. Although anecdotical evidence indicates that LLMs include data relevant to Chile, their capacity to fairly and accurately model Chilean socio-political behavior remains underexplored. Public surveys from Chile and the U.S. that feature comparable political and moral questions are utilized to evaluate and directly compare the models' predictive capabilities across these distinct contexts.

Through this investigation, we aim to assess the extent to which LLMs can accurately capture and replicate socio-political trends in Chile and compare their performance to that in the U.S. context. We examine the role of prompting techniques and the potential for fine-tuning to mitigate performance disparities. Fairness metrics are incorporated to evaluate and compare model performance across various socio-demographic groups to expand this analysis, identifying disparities and their implications. This work not only evaluates the utility of LLMs in underrepresented regions but also addresses broader concerns about fairness, representation, and the ethical use of AI, advancing the discourse on the responsible deployment of LLMs.

The main contributions of this paper are as follows:
\begin{itemize}
    \item A novel framework to measure socio-demographic biases in large language models (LLMs).
    \item A comparative evaluation of LLM predictive capabilities using public surveys from Chile and the U.S.
    \item An analysis of prompting techniques and fine-tuning to mitigate performance disparities.
\end{itemize}

The remainder of this paper is organized as follows: Section 2 details the methodology, including prediction experiments and prompting techniques. Section 3 presents the results, comparing model performance across contexts and socio-demographic groups. Section 4 discusses the findings, highlighting cultural biases and limitations. Section 5 concludes with a summary and future research directions. 


\section{Methodology}
Three prediction experiments based on public surveys have been designed: two concerning political events and one focused on opinion. In the first scenario, the model is tasked with predicting vote choices in presidential elections and referendums. In the second scenario, the model predicts an individual's stance on abortion based on their socio-demographic characteristics. The objective is to evaluate whether LLMs can accurately predict political opinions or voting behavior based on socio-demographic characteristics. The details of the public surveys are as follows:
\begin{itemize}
    \item \textbf{Presidential Election Experiment}: includes both the 2020 U.S. presidential election and the second round of the 2021 Chilean presidential election.
    \item \textbf{Abortion Opinion Experiment}: includes data from both the CEP survey in Chile and the U.S. ANES survey to predict individuals' opinions on abortion based on socio-demographic characteristics. 
    \item \textbf{Constitutional Plebiscite 2022 Experiment}: corresponding to the plebiscite held in Chile to decide on a new constitution.
\end{itemize}

The experiments include different parameters, models, and prompts to optimize the results. The models tested include GPT-3.5 (GPT-3.5-turbo), GPT-4 (GPT-4-turbo), Llama-13b, and Mistral (version 0.2). For cost reasons, GPT-4 is not used for all prompt variants.

\subsection{Datasets}
This research utilizes two public survey datasets, one from Chile and one from the United States. The Chilean dataset comes from the Centro de Estudios Públicos (CEP)~\footnote{\url{https://www.cepchile.cl/opinion-publica/encuesta-cep/}}, which conducts surveys on various socio-political topics. Two CEP surveys were used, with Survey N° 88 (November-December 2022) serving as the test dataset for prediction experiments. Survey N° 86 (April-May 2022)  and N° 89 (N° 89, June-July 2023) were used for fine-tuning, with rows containing missing values for any of the dependent variables or irrelevant columns removed to improve model reliability. Socio-demographic variables include gender, age, region, socio-economic status, educational level, political interest, ideology, and party.

The U.S. dataset is the American National Election Studies (ANES) 2020 Time Series~\footnote{\url{https://electionstudies.org/data-center}}, focusing on electoral behavior and political attitudes. The data initially contained over 8,000 responses, but post-cleaning, it was reduced to 660 rows and 9 columns for comparability with the Chilean dataset. Variables include socio-demographic dimensions similar to those in the CEP data, such as gender, age, region, socio-economic status, educational level, political interest, ideology, and party. The consistent cleaning and selection ensure that the datasets are directly comparable for examining LLM prediction accuracy across different socio-political contexts.

\subsection{\textit{Prompting}}
Research shows that \textit{prompts} significantly influence model performance, with sensitivity varying by task and model~\cite{prompt-prosa,prompt-multichoice,prompt-robust}. Efforts to standardize prompts have led to various design techniques~\cite{prompt-designs}, including Chain-of-Thought prompting~\cite{cot}, which instructs models to explain their reasoning to enhance performance.

The experiments test prompt variations by incorporating event-specific context, adjusting socio-demographic variables, and modifying language or structure to evaluate model performance and sensitivity. Inputs include socio-demographic characteristics, the target question, and response options, with variables such as gender, age, region, urban/rural zone, socio-economic group, education level, religious affiliation, political orientation, party sympathy, and political interest.

The most consistently effective prompt configuration, our default prompt, combines the \textit{chain-of-thought} (CoT) technique, English, few-shot examples~\cite{gpt3} using 5 random examples to enhance clarity and excludes event-specific context. English is selected as the default language due to its optimization in most models, but Spanish is included for its relevance to experiments and supported translation capabilities. Detailed results of the prompt sensitivity analysis can be found in Section~\ref{sec:prompting_appendix} in the Appendix.



To study how socio-demographic variables affect model performance, an ablation analysis is conducted. In this analysis, individual characteristics are systematically removed from the prompt, and the harmonic mean obtained by the most effective LLM for the experiment is recalculated for each condition. Additionally, two alternative prompts are created: one containing only political characteristics and another excluding political variables altogether. This approach enables the examination of the contribution of each variable and assesses the model's reliance on political variables when making predictions. The results are presented in Section~\ref{sec:ablation_table_appendix} in the Appendix.

\subsection{Prediction Limit}
A \textit{Random Forest} is trained \textit{in-sample} to determine the prediction limit on the data. This approach also helps control for differences in the behavior of socio-demographic groups across countries. It is possible that predicting political opinion in Chile is more challenging than in the U.S. due to factors like lower partisanship or greater disinterest. Training the model \textit{in-sample} allows capturing an upper bound on the prediction accuracy, as the model is trained with the same data it will predict.

\subsection{Performance metrics}
To evaluate the performance of LLMs in predicting outcomes, we employ two complementary metrics that address distinct aspects of model accuracy. \textit{Accuracy} focuses on the individual level, assessing how well the model predicts each person's voting behavior. In contrast, the Jensen-Shannon Divergence (\textit{JSS}) operates globally, measuring how closely the predicted overall distribution aligns with the true distribution. 

\subsubsection{\textit{Accuracy}}
Accuracy is a metric that represents the percentage of correctly made predictions. It evaluates the alignment between the model's predictions and the actual data on an individual level (person by person). This metric is particularly relevant for assessing the potential of LLMs in predicting voting outcomes, as it measures how many individual predictions the model gets right. The formula for accuracy is:

$$ 
\text{Accuracy} = \frac{PC}{TP},
$$
where $PC$ represents the number of correct predictions, and $TP$ is the total number of data points to predict.

\subsubsection{Jensen-Shannon \textit{Similarity}}
The Jensen-Shannon Similarity (JSS) quantifies the similarity between two probability distributions, measuring the "distance" or informational difference between them. This metric evaluates whether the model can generate a distribution similar to the original voting distribution, focusing on group-level patterns rather than individual predictions. The formula is:

$$ 
JSS(P \parallel Q) = 1- \frac{1}{2} \big(D_{KL}(P \parallel M) + D_{KL}(Q \parallel M)\big).
$$
Here, $P$ and $Q$ denote the original and predicted probability distributions, respectively, while $M$ is the average of the two distributions, calculated as $M = \frac{1}{2}(P + Q)$. The term $D_{KL}$ represents the Kullback-Leibler Divergence, which measures the divergence between a probability distribution and a reference distribution. 

To compute the total Jensen-Shannon Similarity (JSS) for a model, the data is divided into socio-demographic groups. Each subgroup is then weighted according to its proportion in the dataset, and these weighted values are averaged across all groups to obtain the total.

\subsubsection{Overall Accuracy Equality}
Overall Accuracy Equality is a fairness criterion introduced by \cite{berk2021fairness} that evaluates whether a classifier achieves equal prediction accuracy across protected and unprotected groups. This definition ensures that individuals from each group are equally likely to be correctly classified into their respective classes. 
This metric emphasizes parity in performance across all outcomes and assumes that true negatives are as important as true positives. Moreover, it extends naturally to multi-class classification, requiring that the accuracy for each class is balanced across the groups being compared. 


To address potential sources of bias and disentangle their origins, we adopt in-sample random forest technique to normalize model performance. This normalization allows us to control for variations in predictability across different groups and geographic regions, such as differences in data quality or inherent variability in the target variables. By comparing the relative performance of the primary model against this baseline, we can better identify whether observed disparities stem from biases in the data or limitations in the model itself.

\subsubsection{Logistic Regression}
To deepen the analysis of fairness and better understand potential disparities, we combine the overall accuracy equality criterion with logistic regression, which helps measure how socio-demographic factors influence model prediction accuracy. While Overall Accuracy Equality provides a broad measure of fairness in classification, logistic regression offers insights into how specific variables—such as gender, age, region, and political identity—affect the likelihood of a correct prediction. By modeling prediction accuracy as a binary dependent variable (1 for correct, 0 for incorrect), we can assess how these socio-demographic characteristics influence the fairness of the model's predictions. 

Furthermore, logistic regression allows for the study of the intersection of these socio-demographic groups, enabling the analysis of between-group fairness. This means we can examine how combinations of different characteristics (e.g., age and gender, or political identity and education level) jointly impact model performance, shedding light on whether certain groups experience compounded disadvantages in prediction accuracy.


The logit regression model is specified as:
$$
\log \left( \frac{p}{1-p} \right) = \alpha_j + \beta_1 X_1 + \beta_2 X_2 + \ldots + \beta_n X_n,
$$
where $p$ is the probability of a correct prediction, $\alpha_j$  is the intercept for survey question $j$, and  \( \beta_1, \beta_2, \ldots, \beta_n \) are the coefficients for the socio-demographic variables \( X_1, X_2, \ldots, X_n \).
The coefficients represent the change in the log-odds of a correct prediction for a one-unit change in the corresponding independent variable, holding all other variables constant.

\subsection{\textit{Fine-tuning}}
The impact of Fine-tuning the model with socio-demographic data and survey responses is evaluated to see if it allows the LLM to capture patterns in Chilean society better, improving its performance. The fine-tuning uses data from the CEP surveys, ensuring that the model is not trained on data from the test subjects or survey questions tested. The goal is to train a general model that works across all experiments. Since the training data does not include information from the predicted event, only socio-demographic data and other survey responses are used for training. 

\section{Results}

\subsection{Regional Experiments}
The experiments in Table~\ref{table:performance_metrics} use the default prompt consistently across all models and experiments to enable a realistic comparison of using LLMs as predictors for political events. This choice reflects real-world scenarios where no validation data would be available to optimize prompt performance. 

\begin{table}[t]
\centering
\resizebox{\textwidth}{!}{
\begin{tabular}{|>{\columncolor[HTML]{EFEFEF}}c|c|c|c|c|c|c|c|c|c|c|}
\hline
\cellcolor[HTML]{C0C0C0}{\color[HTML]{000000}}& \multicolumn{4}{|c|}{\cellcolor[HTML]{C0C0C0}{\color[HTML]{000000} Presidential Voting}} & \multicolumn{4}{|c|}{\cellcolor[HTML]{C0C0C0}{\color[HTML]{000000} Abortion}} & \multicolumn{2}{|c|}{\cellcolor[HTML]{C0C0C0}{\color[HTML]{000000} Constitution }} \\ \cline{2-11} 
\cellcolor[HTML]{C0C0C0}{\color[HTML]{000000}Model }& \multicolumn{2}{|c|}{\cellcolor[HTML]{C0C0C0}{\color[HTML]{000000} U.S.}} & \multicolumn{2}{|c|}{\cellcolor[HTML]{C0C0C0}{\color[HTML]{000000} Chile}} & \multicolumn{2}{|c|}{\cellcolor[HTML]{C0C0C0}{\color[HTML]{000000} U.S.}} & \multicolumn{2}{|c|}{\cellcolor[HTML]{C0C0C0}{\color[HTML]{000000} Chile}} & \multicolumn{2}{|c|}{\cellcolor[HTML]{C0C0C0}{\color[HTML]{000000} Referendum}} \\ \cline{2-11} 
\cellcolor[HTML]{C0C0C0}{\color[HTML]{000000} }     & \cellcolor[HTML]{C0C0C0}{\color[HTML]{000000} Acc} & \cellcolor[HTML]{C0C0C0}{\color[HTML]{000000} JSS} & \cellcolor[HTML]{C0C0C0}{\color[HTML]{000000} Acc} & \cellcolor[HTML]{C0C0C0}{\color[HTML]{000000} JSS} & \cellcolor[HTML]{C0C0C0}{\color[HTML]{000000} Acc} & \cellcolor[HTML]{C0C0C0}{\color[HTML]{000000} JSS} & \cellcolor[HTML]{C0C0C0}{\color[HTML]{000000} Acc} & \cellcolor[HTML]{C0C0C0}{\color[HTML]{000000} JSS} & \cellcolor[HTML]{C0C0C0}{\color[HTML]{000000} Acc} & \cellcolor[HTML]{C0C0C0}{\color[HTML]{000000} JSS} \\ \hline
\textit{In-sample Random Forest} & 0.92 & 0.89 & 0.90 & 0.87 & 0.70 & 0.73 & 0.97 & 0.99 & 0.92 & 0.92 \\ \hline
ChatGPT-4 & 0.85 {\footnotesize(0.92)} & 0.84 {\footnotesize(0.94)} & 0.62 {\footnotesize(0.69)} & 0.86 {\footnotesize(0.99)} & \textbf{0.61} {\footnotesize(0.87)}& 0.69 {\footnotesize(0.95)} & \textbf{0.52} {\footnotesize(0.54)}& 0.76 {\footnotesize(0.77)} & \textbf{0.67} {\footnotesize(0.73)}& 0.89 {\footnotesize(0.97)} \\ \hline
ChatGPT-3.5 & 0.87 {\footnotesize(0.95)} & \textbf{0.93} {\footnotesize(1.04)}& 0.64 {\footnotesize(0.71)} & \textbf{0.92} {\footnotesize(1.06)}& 0.56 {\footnotesize(0.80)} & 0.79 {\footnotesize(1.08)} & 0.43 {\footnotesize(0.44)} & 0.70 {\footnotesize(0.71)} & 0.58 {\footnotesize(0.63)} & 0.75 {\footnotesize(0.82)} \\ \hline
Llama-13b & \textbf{0.88} {\footnotesize(0.96)}& \textbf{0.93} {\footnotesize(1.04)}& 0.69 {\footnotesize(0.77)} & 0.87 {\footnotesize(1.00)} & \textbf{0.61} {\footnotesize(0.87)}& \textbf{0.89} {\footnotesize(1.22)}& \textbf{0.52} {\footnotesize(0.54)}& 0.84 {\footnotesize(0.85)} & 0.65 {\footnotesize(0.71)} & \textbf{0.90} {\footnotesize(0.98)}\\ \hline
Mistral & 0.84 {\footnotesize(0.91)} & 0.83 {\footnotesize(0.93)} & 0.54 {\footnotesize(0.60)} & 0.78 {\footnotesize(0.90)} & 0.59 {\footnotesize(0.84)} & 0.84 {\footnotesize(1.15)} & 0.45 {\footnotesize(0.46)} & 0.83 {\footnotesize(0.84)} & 0.47 {\footnotesize(0.51)} & 0.67 {\footnotesize(0.73)} \\ \hline
\textit{Fine-tuned} (F2) & - & - & \textbf{0.71} {\footnotesize(0.79)}& 0.75 {\footnotesize(0.86)} & - & - & 0.50 {\footnotesize(0.52)} & \textbf{0.88} {\footnotesize(0.89)}& 0.63 {\footnotesize(0.68)} & 0.83 {\footnotesize(0.90)} \\ \hline
\end{tabular}
}
\caption{Performance metrics across different models for various experiments, using the same \textit{chain of thought} prompt across all tests. Each cell contains accuracy (Acc) and JSS values, with the relative ratio to the \textit{In-sample Random Forest} in parentheses.}
\label{table:performance_metrics}
\end{table}


\subsubsection{Country Comparison} The results in Table~\ref{table:performance_metrics} show that the evaluation of models using context from the United States and Chile show significant differences, which may be due to cultural and dataset-specific biases in large language models. In general, models perform better on U.S. datasets than on those from Chile, indicating the predominant U.S.-centric focus of many pre-training corpora. For instance, the accuracy (Acc) and Jensen-Shannon Similarity Score (JSS) for presidential voting in the U.S. are consistently higher than those for Chile across all models, especially in terms of accuracy. This trend is also observed in the abortion question.

Even well-calibrated models, such as ChatGPT~\cite{kadavath2022language,tian2023just}, struggle to achieve the same level of accuracy in the Chilean context as they do in the U.S.. To ensure a fair comparison across countries, we normalize the performance of each model using the in-sample random forest to account for differences in predictability and dataset characteristics between the U.S. and Chile. The relative performance is displayed in parentheses in Table~\ref{table:performance_metrics}.  The consistent performance gap, particularly in terms of relative accuracy, suggests that cultural and dataset-specific biases are influencing the large language models. 
Those trained primarily on U.S.-centric data struggle to interpret Chilean datasets effectively, likely due to differences in language use, cultural context, and data structure.

The F2 Llama model, which has been fine-tuned for the Chilean context, shows only minor changes. However, the analysis of presidential voting suggests that there is potential for better adapting language models to local contexts. This adaptation can help address limitations and improve their effectiveness in various settings.

\subsubsection{Model Level Comparison} The comparative performance of different models in Table~\ref{table:performance_metrics} reveals distinct strengths and weaknesses that are influenced by their architectures and pre-training data. Among the models, \textbf{Llama-13B emerges as the best performer}, consistently achieving the highest accuracy and balanced results across tasks and countries. It frequently outperforms other models and surpasses the in-sample Random Forest benchmark in critical tasks such as the U.S. Abortion and Chile Referendum predictions, which makes it particularly good for population-level predictions. For additional insights, Appendix Figure~\ref{table:ablation_table_appendix} provides a detailed ablation study of Llama-13B’s performance, varying the socio-demographic features used for prediction.

ChatGPT-4 also demonstrates robust performance, delivering strong results across most tasks and metrics, as shown in Table~\ref{table:performance_metrics}. In contrast, ChatGPT-3.5 exhibits greater variability; while its JSS ratios remain competitive across several datasets, its accuracy falls short in tasks such as Chile's Abortion and the referendum. Mistral, on the other hand, struggles to adapt to non-U.S. data, particularly in the Chilean datasets, where it shows relatively low accuracy for the Abortion and Referendum tasks, reflecting its limitations in generalizing across different socio-political contexts.

Interestingly, many LLMs, including ChatGPT-4 and Llama-13B, outperform the in-sample Random Forest in Jensen-Shannon Similarity (JSS), demonstrating their capacity to reflect population-level patterns in political behavior. This suggests that these models excel at capturing broader macro-level trends rather than focusing solely on individual-level prediction accuracy, making them valuable tools for societal-level analyses and understanding their limitations in predicting individual opinions.

Notably, the fine-tuned model does not outperform its original, pre-trained counterpart. Its lower accuracy and JSS compared to Llama-13B suggest that fine-tuning introduces a bias toward specific options, leading to distributions that deviate from the original data. A notable exception is observed in the Abortion experiment, where the fine-tuned model achieves a JSS score near 0.9, closely resembling the original data distribution. Nonetheless, this remains an outlier rather than a consistent trend, which reinforces the overall superiority of Llama-13B for these tasks. It also raises the possibility that other fine-tuning techniques and datasets might enhance the model.


\begin{figure}[t!]
    \centering
    \begin{minipage}[b]{0.65\textwidth}
        \centering
        \includegraphics[width=\textwidth]{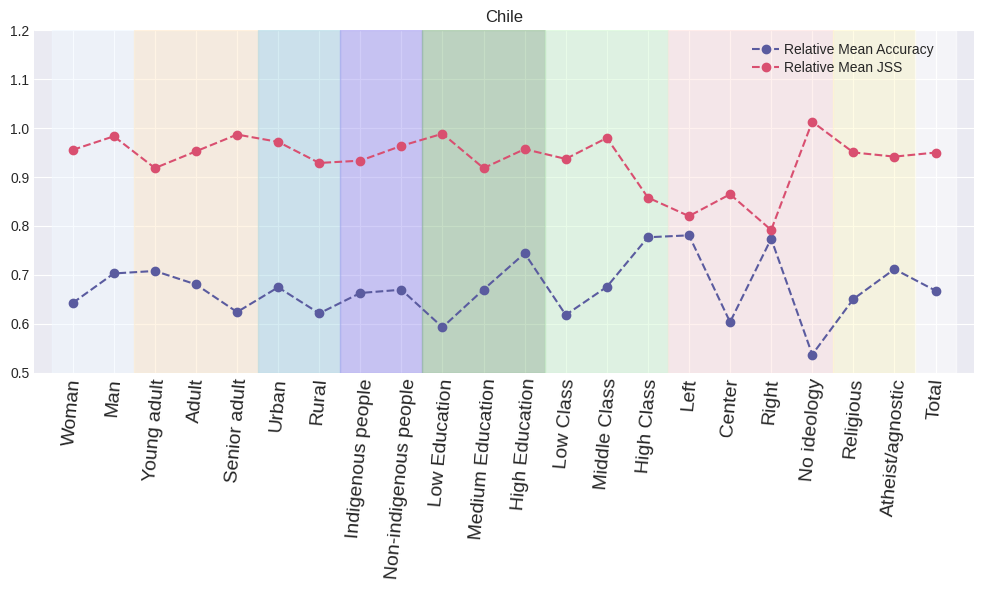}
      \end{minipage} 
      \begin{minipage}[b]{0.65\textwidth}
        \centering
        \vspace{0cm} 
        \includegraphics[width=\textwidth]{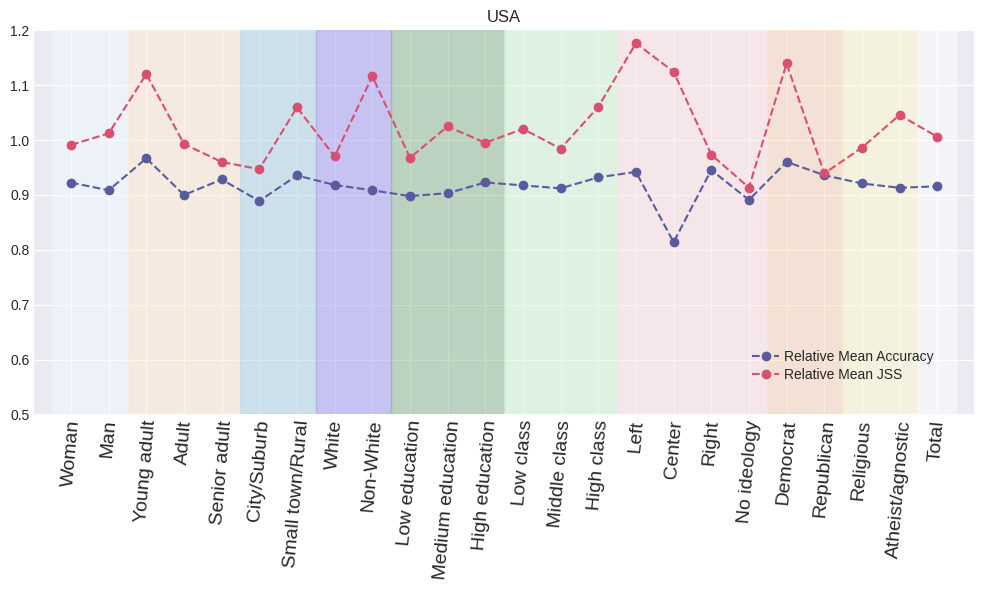}
      \end{minipage}
    \caption{\textbf{Relative Performance Metrics (Accuracy and Jensen-Shannon Similarity) for the Llama-13B Model Compared to the In-Sample Random Forest.} This figure shows the relative mean Accuracy and Jensen-Shannon Similarity (JSS) for each socio-demographic group across multiple experiments conducted in Chile (left figure) and the United States (right figure). Results are normalized by the performance of the in-sample Random Forest model, ensuring consistent comparisons across groups and regions. The analysis uses identical models and prompts for all experiments.}
    \label{fig:linea_promedio}
\end{figure}

\subsection{Socio-Demographic Experiments}
For the following analyses, we will focus on the overall best-performing model, which is the pre-trained Llama-13B. For socio-demographic results related to the other models, please refer to Appendix Figure~\ref{fig:linea_promedio_appendix}. However, it is worth noting that most trends are consistent across the models. Specifically, we will examine potential biases in model performance across socio-demographic groups for the three survey questions from Chile and the two survey questions from the United States.

Figure~\ref{fig:linea_promedio} displays the relative mean accuracy and relative mean Jensen-Shannon Similarity (JSS) for various socio-demographic groups in Chile and the U.S.. The performance metrics are normalized by the in-sample random forest, allowing for a comparative analysis across different groups.  The Random Forest model acts as an upper bound on the predictability of the data, providing a benchmark for evaluating the performance of the Llama model. Refer to Figure~\ref{fig:linea_promedio_appendix} in the Appendix for the detailed values for each model. 
The analysis will focus on identifying potential model biases. The analysis reveals that Llama model exhibits accuracy bias more pronounced in women, groups with lower education levels, lower social classes, older adults and political ideologies. 

\paragraph{Gender Disparities}
In Chile, the Llama model exhibits notable gender disparities, with higher accuracy and Jensen-Shannon Similarity (JSS) for men compared to women. This performance gap, contrasted with the consistent performance of the Random Forest model, strongly suggests model bias against women in the Chilean context (see Appendix Figure~\ref{fig:linea_promedio_appendix}). In the United States, however, the performance across genders remains stable, with women showing slightly better JSS performance in both the Random Forest and Llama models, indicating an absence of gender-related bias.

\paragraph{Age Disparities}
When examining age groups, both Chile and the U.S. display a decline in model accuracy as the age of individuals increases. This drop is more pronounced in Chile, whereas in the U.S., the variability is less extreme, with the model showing a modest decline in accuracy for adults but a smaller decrease for seniors compared to young adults. Interestingly, JSS trends diverge between the two countries. While accuracy decreases for older individuals in Chile, the predicted distributions (as measured by JSS) become more aligned with the overall distribution. In the U.S., JSS declines consistently with age, mirroring the drop in accuracy.

\paragraph{Education and Income Level Disparities}
Education level also plays a significant role in model performance. In both countries, the models are less accurate for individuals with lower education levels, although this effect is much less pronounced in the U.S.. A similar pattern is observed for social class in Chile, where accuracy decreases for lower-income classes, whereas in the U.S., accuracy remains relatively consistent across classes.

\paragraph{Political ideology Disparities}
Political ideology reveals distinct patterns. In both Chile and the U.S., individuals with left- or right-leaning ideologies achieve significantly higher relative mean accuracy compared to those with centrist views, even though we account for the fact that center-leaning individuals are more unpredictable (see Random Forest in Appendix Figure~\ref{fig:linea_promedio_appendix}).  This suggests that the model is better attuned to the perspectives of individuals with strong political views. In Chile, individuals without a stated ideology are less predictable, yet their predicted distributions align closely with the overall distribution, as indicated by their high JSS scores. This shows the nuanced aspect of the model's performance, where consistency in overall predictions does not necessarily correlate with accuracy.

\begin{table}[ht]
\centering
\scalebox{0.9}{
\begin{tabular}{lcccc}
\toprule
\textbf{Chile}  \\
\midrule
\textbf{Socio-demographic} & \textbf{Model 1} & \textbf{Model 2} & \textbf{Model 3} & \textbf{Model 4} \\
\midrule
\textbf{Question} & & & & \\
Abortion & 0.45 (0.15) & 0.25 (0.45) & 0.028 (0.943) & -0.043 (0.905) \\
Referendum & 1.03 (0.001)*** & 0.83 (0.011)** & 0.615 (0.125) & 0.539 (0.132) \\
Presidential & 1.2 (0.000)*** & 1.04 (0.001)*** & 0.825 (0.040)** & 0.749 (0.037)** \\
\midrule
\textbf{Gender (ref = Man)} & & & & \\
Woman & -0.24 (0.017)** & 0.10 (0.58) & -0.266 (0.010)** & -0.251 (0.014)** \\
\midrule
\textbf{Age (ref = Young Adult)} & & & & \\
Adult & -0.19 (0.25) & -0.18 (0.27) & -0.159 (0.341) & -0.168 (0.309) \\
Senior Adult & -0.38 (0.04)** & -0.35 (0.06)* & -0.347 (0.063)* & -0.342 (0.064)* \\
\midrule
\textbf{Region (ref = Urban)} & & & & \\
Rural & -0.04 (0.80) & -0.02 (0.91) & -0.041 (0.785) & -0.023 (0.879) \\
\midrule
\textbf{Indigenous (ref = Non-indigenous)} & & & & \\
Indigenous & 0.11 (0.56) & 0.10 (0.59) & 0.123 (0.494) & 0.117 (0.514) \\
\midrule
\textbf{Education (ref = High)} & & & & \\
Low & -0.51 (0.001)*** & -0.53 (0.000)*** & -0.103 (0.750) & -0.549 (0.000)*** \\
Medium & -0.36 (0.004)** & -0.36 (0.004)** & 0.248 (0.452) & -0.377 (0.003)** \\
\midrule
\textbf{Income (ref = High)} & & & & \\
Low & -0.01 (0.97) & 0.02 (0.95) & -0.052 (0.846) & 0.072 (0.784) \\
Middle & -0.07 (0.78) & -0.05 (0.826) & -0.091 (0.707) & -0.036 (0.878) \\
\midrule
\textbf{Political Identity (ref = None)} & & & & \\
Left & 1.09 (0.000)*** & 1.093 (0.000)*** & 1.913 (0.000)*** & 1.801 (0.000)*** \\
Center & 0.11 (0.426) & 0.094 (0.49) & 0.440 (0.153) & 0.619 (0.010)** \\
Right & 1.15 (0.000)*** & 1.132 (0.000)*** & 1.645 (0.000)*** & 1.424 (0.000)*** \\
\midrule
\textbf{Religion (ref = Not Religious)} & & & & \\
Religious & -0.23 (0.03)** & 0.06 (0.70) & -0.224 (0.040)** & 0.394 (0.108) \\
\midrule
\textbf{Interaction Terms} & & & & \\
\midrule
Woman $\times$ Religious & - & -0.52 (0.015)** & - & - \\
\midrule
Low Education $\times$ Left & - & - & -1.394 (0.002)*** & - \\
Medium Education $\times$ Left & - & - & -0.888 (0.045)** & - \\
Low Education $\times$ Center & - & - & 0.034 (0.928) & - \\
Medium Education $\times$ Center & - & - & -0.694 (0.062)* & - \\
Low Education $\times$ Right & - & - & -0.692 (0.135) & - \\
Medium Education $\times$ Right & - & - & -0.524 (0.266) & - \\
\midrule
Left $\times$ Religious & - & - & - & -1.087 (0.002)*** \\
Center $\times$ Religious & - & - & - & -0.732 (0.011)** \\
Right $\times$ Religious & - & - & - & -0.423 (0.279) \\
\bottomrule
\end{tabular}
}
\caption{Logistic Regression Results for Correct Prediction of Individual Answers in Chile. Standard errors are displayed in parentheses. ***p < 0.01; **p < 0.05; *p < 0.1.}
\label{tab:logit_regression_cl}
\end{table}

\begin{table}[ht]
\centering
\scalebox{0.9}{
\begin{tabular}{lccc}
\toprule
\textbf{United States} & \textbf{Model 1} & \textbf{Model 2} & \textbf{Model 3} \\
\midrule
\textbf{Socio-demographic Group} & \textbf{Coefficient (p-value)} & \textbf{Coefficient (p-value)} & \textbf{Coefficient (p-value)} \\
\midrule
\textbf{Question} & & & \\
Abortion & 0.54 (0.466) & 0.891 (0.264) & 0.451 (0.543) \\
Presidential & 2.24 (0.003)*** & 2.632 (0.001)*** & 2.137 (0.004)*** \\
\midrule
\textbf{Gender (ref = Man)} & & & \\
Woman & 0.01 (0.960) & -0.758 (0.078)* & 0.150 (0.367) \\
\midrule
\textbf{Age (ref = Young Adult)} & & & \\
Adult & -0.41 (0.222) & -0.374 (0.267) & -0.399 (0.234) \\
Senior Adult & 0.11 (0.756) & 0.138 (0.691) & 0.114 (0.742) \\
\midrule
\textbf{Region (ref = Urban)} & & & \\
Rural & 0.20 (0.200) & 0.220 (0.150) & 0.194 (0.201) \\
\midrule
\textbf{Race (ref = White)} & & & \\
Non-White & 0.12 (0.514) & 0.116 (0.516) & 0.443 (0.091)* \\
\midrule
\textbf{Education (ref = High)} & & & \\
Low & -0.13 (0.870) & -0.105 (0.893) & -0.055 (0.943) \\
Medium & -0.02 (0.980) & 0.042 (0.946) & 0.021 (0.972) \\
High & & 0.146 (0.813) & 0.116 (0.848) \\
\midrule
\textbf{Income (ref = High)} & & & \\
Low & -0.50 (0.032)** & -0.496 (0.031)** & -0.497 (0.030)** \\
Middle & -0.26 (0.212) & -0.278 (0.179) & -0.269 (0.193) \\
\midrule
\textbf{Political Identity (ref = None)} & & & \\
Left & 1.08 (0.000)*** & 0.189 (0.640) & 1.069 (0.000)*** \\
Center & -0.49 (0.054)* & -0.789 (0.049)** & -0.514 (0.046)** \\
Right & 0.24 (0.341) & -0.209 (0.583) & 0.223 (0.370) \\
\midrule
\textbf{Religion (ref = Not Religious)} & & & \\
Religious & -0.23 (0.128) & -0.218 (0.157) & -0.236 (0.124) \\
\midrule
\textbf{Interaction Terms} & & & \\
\midrule
Left × Woman & & 1.578 (0.003)*** & \\
Center × Woman & & 0.429 (0.403) & \\
Right × Woman & & 0.728 (0.136) & \\
\midrule
Non-White × Woman & & & -0.590 (0.081)* \\
\bottomrule
\end{tabular}
}
\caption{Logistic Regression Results for Correct Prediction of Individual Answers in the U.S. (Models 2 and 3). Standard errors are displayed in parentheses. ***p < 0.01; **p < 0.05; *p < 0.1.}
\label{tab:logit_regression_us}
\end{table}

\subsection{Intersectional Subgroup Fairness}
Examining fairness solely through individual demographic attributes, such as gender or race, overlooks the compounded discrimination experienced by individuals with intersecting marginalized identities~\cite{kearns2018preventing,kearns2019empirical,wang2022towards}. For instance, a model that performs equitably across gender and race categories independently may still exhibit significant unfairness toward specific subgroups, such as Black women. 

To empirically evaluate intersectional fairness, the overall accuracy equality criterion used in the previous section can be extended to pairs of socio-demographic groups. Visual representations of this analysis are provided in Appendix Figures~\ref{fig:sociodemographic_matrix} and~\ref{matriz_acc_eeuu_eng}. However, addressing fairness across more than two intersecting groups or accounting for data imbalances across the intersection of groups requires a more robust analytical approach. For this purpose, we propose leveraging logistic regression.

The logistic regression analysis offers deeper insights into how socio-demographic factors influence model accuracy across three experiments conducted in Chile. Significant disparities emerge along the lines of gender, education level, and political identity, suggesting that the model’s predictions may exhibit systematic biases. Tables~\ref{tab:logit_regression_cl} and~\ref{tab:logit_regression_us} summarize the results of the logistic regression, where the dependent variable indicates whether the model accurately predicted an individual’s response. The regression coefficients quantify the impact of socio-demographic variables on the likelihood of a correct prediction, allowing us to identify specific groups that may be disproportionately disadvantaged. This approach enables a nuanced examination of fairness, particularly for groups at the intersection of multiple socio-demographic identities.

\subsubsection{Chile}
There is evidence of gender and age-related biases in Chile, with the model performing less accurately for women and senior adults. Education level also significantly impacts accuracy, with lower education levels associated with inferior accuracy. Political identity has a strong influence on prediction accuracy; particularly, left and right identities are associated with higher accuracy. Religious affiliation is associated with lower accuracy. People's region and income, if they are Indigenous, do not significantly affect prediction accuracy.

Examining all the negative coefficients suggests that certain intersections of groups are less likely to have their responses accurately predicted by the model. In particular, senior religious women with low education levels represent the intersectional subgroup with the lowest prediction accuracy. This indicates that the model's prediction accuracy varies among different socio-demographic groups, which raises concerns about fairness and representation for these individuals.

Analyzing the interaction coefficients in a logistic regression model provides insights into how the combination of different socio-demographic factors influences model accuracy. Interaction coefficients show whether the combined effect of two or more factors differs from the sum of their individual effects, offering a deeper understanding of how these factors work together. In our extended logistic regression analysis (Table~\ref{tab:logit_regression_cl}), we focus on models where interaction terms are statistically significant.

Notably, Model 2 reveals a significant negative interaction between being a woman and identifying as religious, which results in a more negative response than the individual effects of gender and religion alone. Similarly, the interactions between Education and Political Identity in Model 3 indicate that individuals with lower or medium education who identify as left-wing exhibit a stronger negative response to the dependent variable than expected from the individual effects of education or political identity. Furthermore, the Religious and Political Identity interaction in Model 4 shows that religious individuals with left-wing or center political views show a significantly lower likelihood of accurate prediction compared to those without religious affiliation, demonstrating how religious beliefs can influence and modify political perspectives.

\subsubsection{United States}
Model 1 in Table~\ref{tab:logit_regression_us} does not exhibit significant gender or age-related biases in the United States, as the accuracy remains consistent across these groups. Education level does not significantly impact prediction accuracy. However, political identity, particularly left-leaning views, is associated with higher accuracy. Religious affiliation does not significantly affect prediction accuracy. Factors such as region and race also do not substantially influence the model's performance. However, income level does have a notable impact, with the model performing less accurately for individuals with low income. 
The intersection of groups with the most negative bias is low-income adults with center-leaning views. 


When examining the interaction terms, significant variations in prediction accuracy based on socio-demographic intersections emerge. In Model 2, the interaction between left-leaning political identity and being a woman significantly improves prediction accuracy, indicating that left-wing women are more accurately predicted than other groups. This pattern does not hold for women with center or right-leaning political views. Interestingly, in Model 3, race becomes a significant factor: the interaction between being non-white and a woman shows a negative effect, suggesting a potential bias in the model's prediction for non-white women compared to other groups. This highlights a contrast between the first and third models, with race emerging as a more relevant factor in Model 3, especially in the context of gender.

\section{Discussion}

The comparison between the United States and Chile reveals performance gaps that originate from cultural and pre-training dataset-specific biases affecting large language models. Models, including ChatGPT and Llama-13B, consistently perform better on datasets from the U.S. compared to those from Chile, as demonstrated by higher accuracy and Jensen-Shannon Similarity Scores (JSS) when predicting presidential voting and abortion-related questions in the U.S. The disparity continued even after controlling for dataset predictability and examining biases among socio-demographic groups. These results reveal the U.S.-centric nature of the pre-training corpora, which provides models with a deeper understanding of U.S.-specific contexts while leaving them less effective at interpreting Chilean data.

In terms of evaluation metrics, our results show that while the models excel in JSS, making them more reliable for predicting group-level trends and distributions, they are less accurate in predicting individual outcomes. This suggests that, while the models are adept at capturing broad population patterns, they struggle to account for the nuances of individual socio-political identities.
This gap presents an opportunity to better capture socio-political dynamics such as the decline of partisanship in Chile~\cite{delacerdaUnstableIdentitiesDecline2022}, shifts in party identification~\cite{bargstedPartyIdentificationEncapsulated2018}, and partisan polarization~\cite{jacobsonPartisanPolarizationAmerican2013,westPartisanshipSocialIdentity2020}. Thus, providing essential context for assessing the models' effectiveness in capturing political phenomena and interpreting their biases.

\subsection{Socio-Demographic Biases}
Our analysis of socio-demographic biases reveals distinct patterns between the United States and Chile. In the United States, political identity, especially left-leaning views, along with low income, significantly impacts prediction accuracy. Notably, there are substantial interactions between political identity, gender, and race. Among these groups, women with left-leaning views are particularly negatively affected. Additionally, the model reveals a pronounced racial bias that especially impacts non-white women,  showing that the intersection of race and gender influences predictive outcomes. 

In Chile, the socio-demographic biases differ, with particular biases affecting women, older adults, religious individuals, and those with lower educational levels.  
With regard to political identity, the model shows a more nuanced influence on prediction accuracy, with center-leaning groups being less well-represented. The interaction between these factors, such as the combined effect of being a woman and identifying as religious, further exacerbates prediction inaccuracies.

The interaction between political identity and gender, while relevant, does not exhibit the same level of significance as in the U.S. models. In contrast, the interactions between education and political identity indicate that individuals with lower or medium education who identify as left-leaning or center-leaning exhibit a stronger negative response to the dependent variable than expected from the individual effects of education or political identity. Similarly,  the religious and political identity interaction shows that religious individuals with left-wing or center political views show a negative bias prediction compared to those without religious affiliation.


\subsection{Future Work}

In future research, one promising direction is exploring advanced fine-tuning techniques, such as Direct Preference Optimization (DPO)~\cite{rafailov2024direct}, which has shown potential for improving model alignment with human preferences. DPO can help mitigate some of the biases observed in this study by more effectively allowing models to better reflect socio-political opinions. 

Moreover, incorporating more diverse and representative datasets during the training and fine-tuning phases would enhance the models' ability to capture a wider range of socio-political identities. This would help create models that are not only more accurate in predicting individual behaviors but also more aligned with real-world demographic diversity.

The generalization of model performance across different countries or regions is limited by the cultural and political differences that may not be well-represented in the pre-training datasets. For example, a model that performs well in the United States may not necessarily transfer effectively to Chile or other countries with distinct political landscapes, as observed in our findings. Cross-cultural adaptability remains a significant challenge for large language models, particularly when predicting behavior based on region-specific socio-political factors.

Future work should address the limited generalization of model performance across regions with distinct cultural and political landscapes. As observed in our findings, models that perform well in one country may not transfer effectively to others, highlighting the challenge of cross-cultural adaptability in predicting behavior influenced by region-specific socio-political factors.

\subsection{Limitations}
While our study provides valuable insights into the biases and disparities in large language models, there are several limitations to consider. One key limitation is the potential selection bias in survey responses, which may skew the data towards certain demographic groups, such as those with higher internet access or particular political leanings. These biases in the survey sample could affect the accuracy of the model's predictions, particularly in regions or groups underrepresented in the data.

\section{Conclusion}
This study highlights significant disparities in large language model performance, influenced by cultural contexts and socio-demographic factors embedded in training data. While models are more effective at predicting group-level distributions (as evidenced by JSS performance), they remain less reliable at capturing individual-level behavior. 

To address these limitations, context-specific strategies are essential. This could involve collecting more diverse and representative training data, incorporating socio-demographic characteristics during fine-tuning, and applying fairness constraints to mitigate biases. Future research should focus on building models that better capture the nuances of political identity and behavior, ensuring more equitable performance across different populations.

The disparities and biases identified in this study emphasize the need to customize models to fit specific cultural and socio-political contexts. Our findings indicate that the socio-demographic biases present in these models are unique, implying that addressing them requires tailored strategies that consider the socio-political environments in which the models are utilized. By closing these gaps, we can enhance the reliability and fairness of predictive models, making them more effective tools for understanding human behavior across diverse global settings.

\section{Ethical Considerations Statement}
This study exclusively utilized publicly available datasets to evaluate the performance of large language models, ensuring compliance with ethical research standards and privacy protection. No sensitive or personally identifiable information was used. The research prioritized beneficence by seeking to identify and mitigate socio-demographic and cultural biases in AI models, aiming to enhance fairness and inclusivity, fostering accountability and responsible AI development.

\section{Adverse Impact Statement}
This research highlights the performance disparities and biases in large language models, particularly regarding socio-demographic and cultural factors. While the goal is to improve model fairness, the findings may inadvertently reinforce existing stereotypes or perpetuate biases if not carefully addressed in future model development and deployment. The reliance on publicly available datasets, while ethically sound, might limit the generalizability of the conclusions to broader, more diverse populations.

\bibliographystyle{ACM-Reference-Format}
\bibliography{bib}

\newpage
\appendix

\section{Appendix}

\subsection{Model Comparison}
\begin{figure}[h]
    \centering
    \begin{minipage}[b]{0.49\textwidth}
        \centering
        \includegraphics[width=\textwidth]{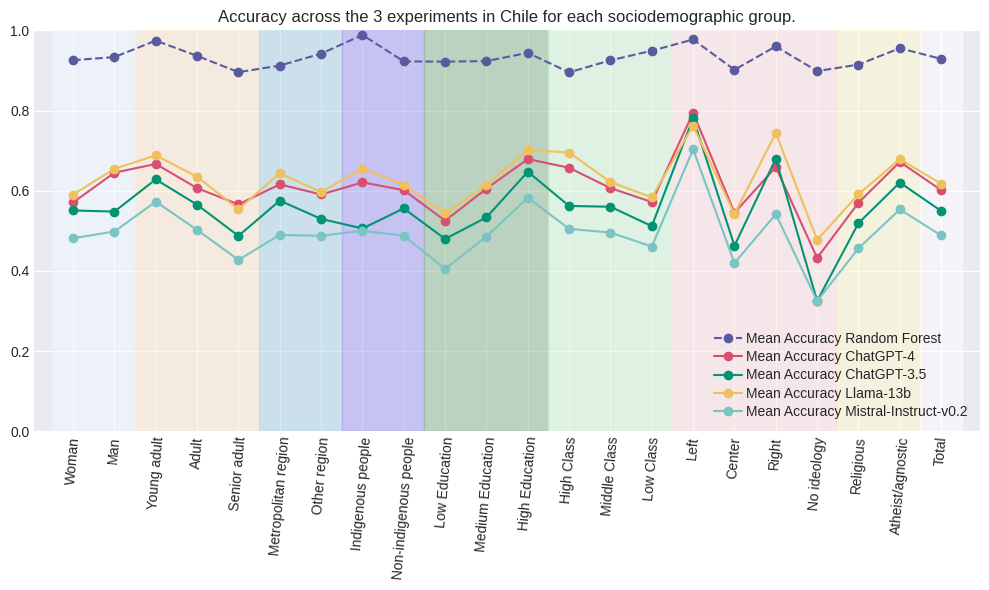}
      \end{minipage} 
      \begin{minipage}[b]{0.49\textwidth}
        \centering
        \vspace{0cm} 
        \includegraphics[width=\textwidth]{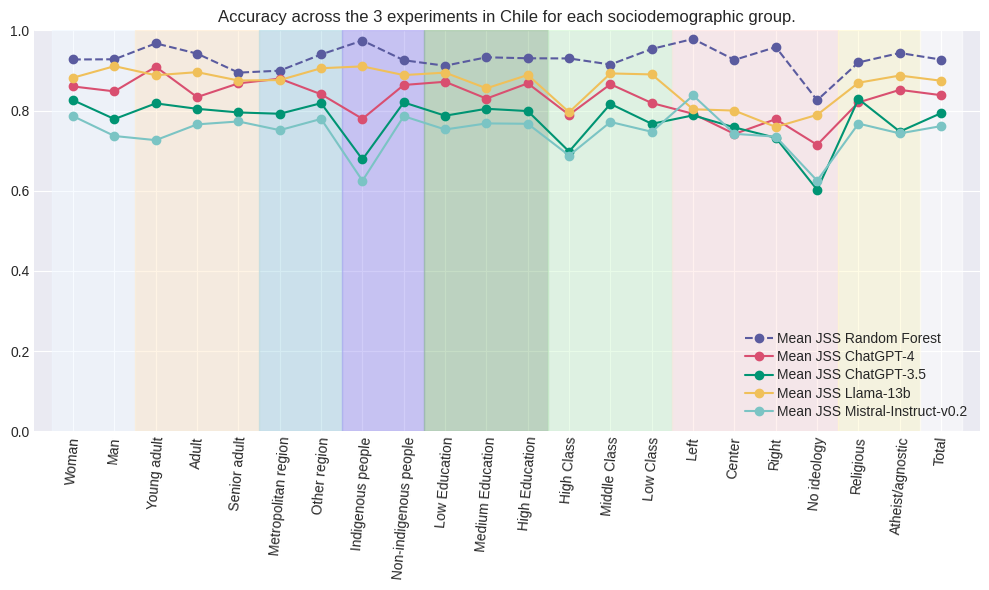}
      \end{minipage}
    \begin{minipage}[b]{0.49\textwidth}
        \centering
        \includegraphics[width=\textwidth]{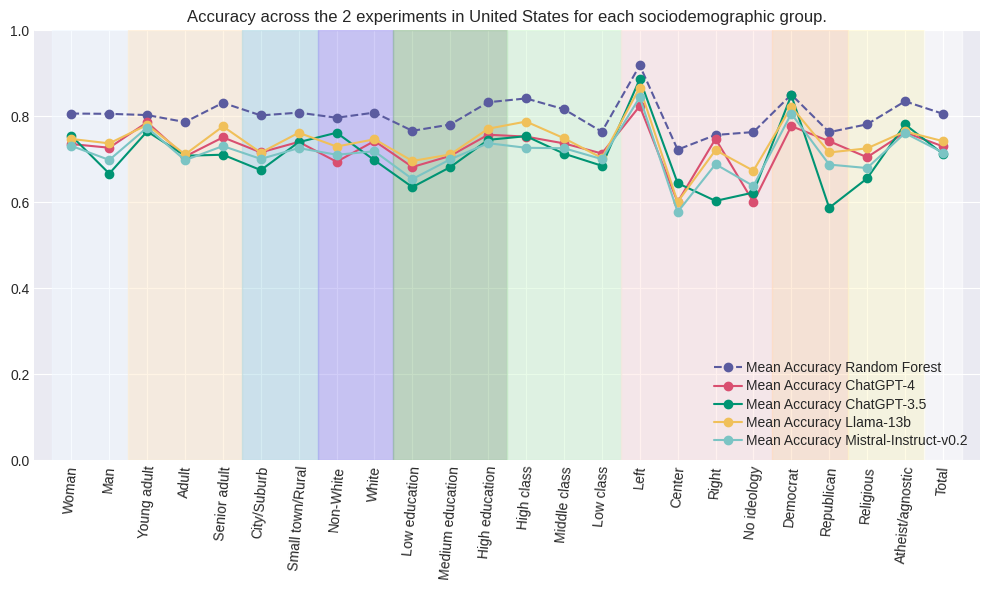}
      \end{minipage} 
      \begin{minipage}[b]{0.49\textwidth}
        \centering
        \vspace{0cm} 
        \includegraphics[width=\textwidth]{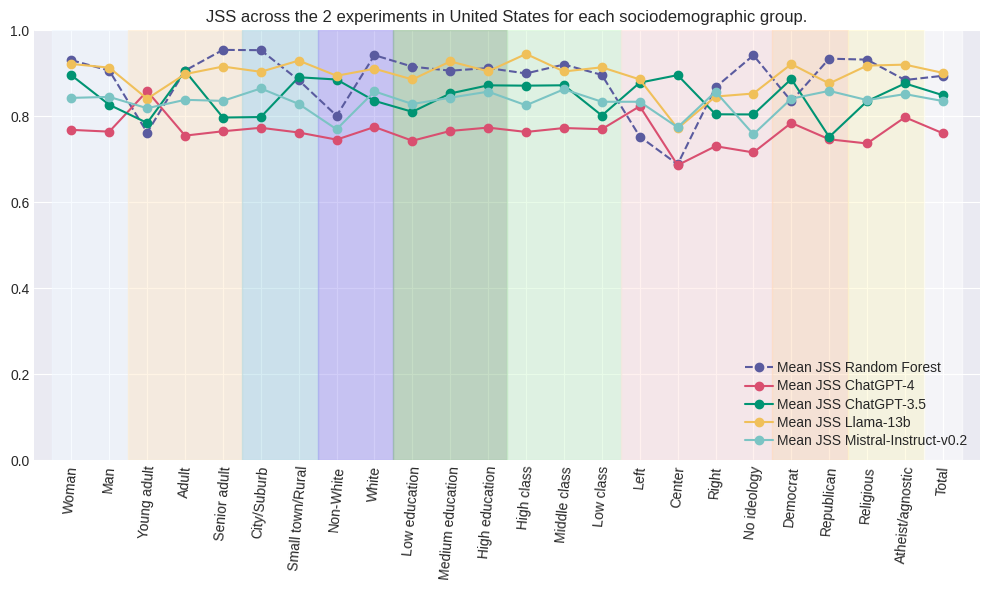}
      \end{minipage}
    \caption{\textbf{Mean Accuracy and Jensen-Shannon Similarity (JSS) across Socio-Demographic Groups for all Models.} The figure shows the mean Accuracy and JSS for each socio-demographic group across multiple experiments conducted in Chile (three survey questions) and the United States (two survey questions) using the same models and prompts.}
    \label{fig:linea_promedio_appendix}
\end{figure}

\begin{figure}[h] 
    \centering 
    \includegraphics[scale=0.45]{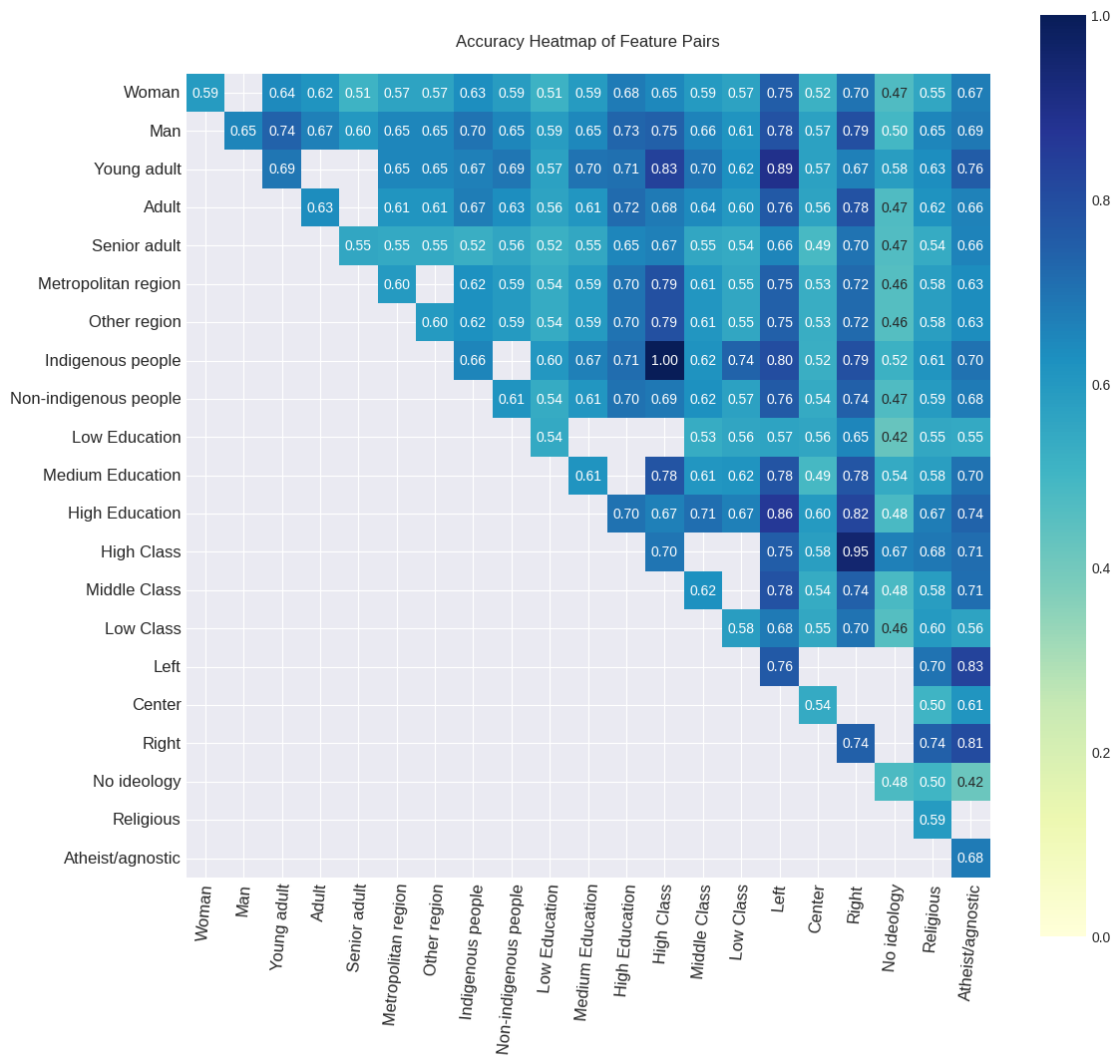} 
     \caption{ Mean accuracy by pairs of socio-demographic groups in Chile. The matrix shows for each pair of socio-demographic groups the average accuracy obtained between the 3 experiments using the same model and \textit{prompts}.} \label{fig:sociodemographic_matrix} 
\end{figure}

\subsection{Socio-Demographic Experiments}

The results in Figure~\ref{fig:sociodemographic_matrix} reveal significant disparities in model accuracy across intersections of demographic subgroups. For gender and age, men consistently receive more accurate predictions than women across all age categories: young adults (0.74 for men vs. 0.64 for women), adults (0.67 for men vs. 0.62 for women), and senior adults (0.60 for men vs. 0.51 for women). Regionally, men in both metropolitan and other regions have higher accuracy (0.65) compared to women (0.57). In terms of Indigenous status, Indigenous men (0.70) and non-indigenous men (0.65) both have higher accuracy than their female counterparts (0.63 and 0.59, respectively). Education level also impacts accuracy, with higher education corresponding to better performance for both genders, but men still outperform women in both low (0.62 for men vs. 0.58 for women) and high education groups (0.75 for men vs. 0.65 for women). Political ideology shows similar trends, with men having higher accuracy across left (0.79 for men vs. 0.70 for women), center (0.50 for men vs. 0.47 for women), and right (0.65 for men vs. 0.55 for women) affiliations. These findings highlight a pervasive gender bias in the model's predictions that needs targeted interventions to ensure fairness across all subgroups.

\begin{figure}[h] 
    \centering 
    \includegraphics[scale=0.45]{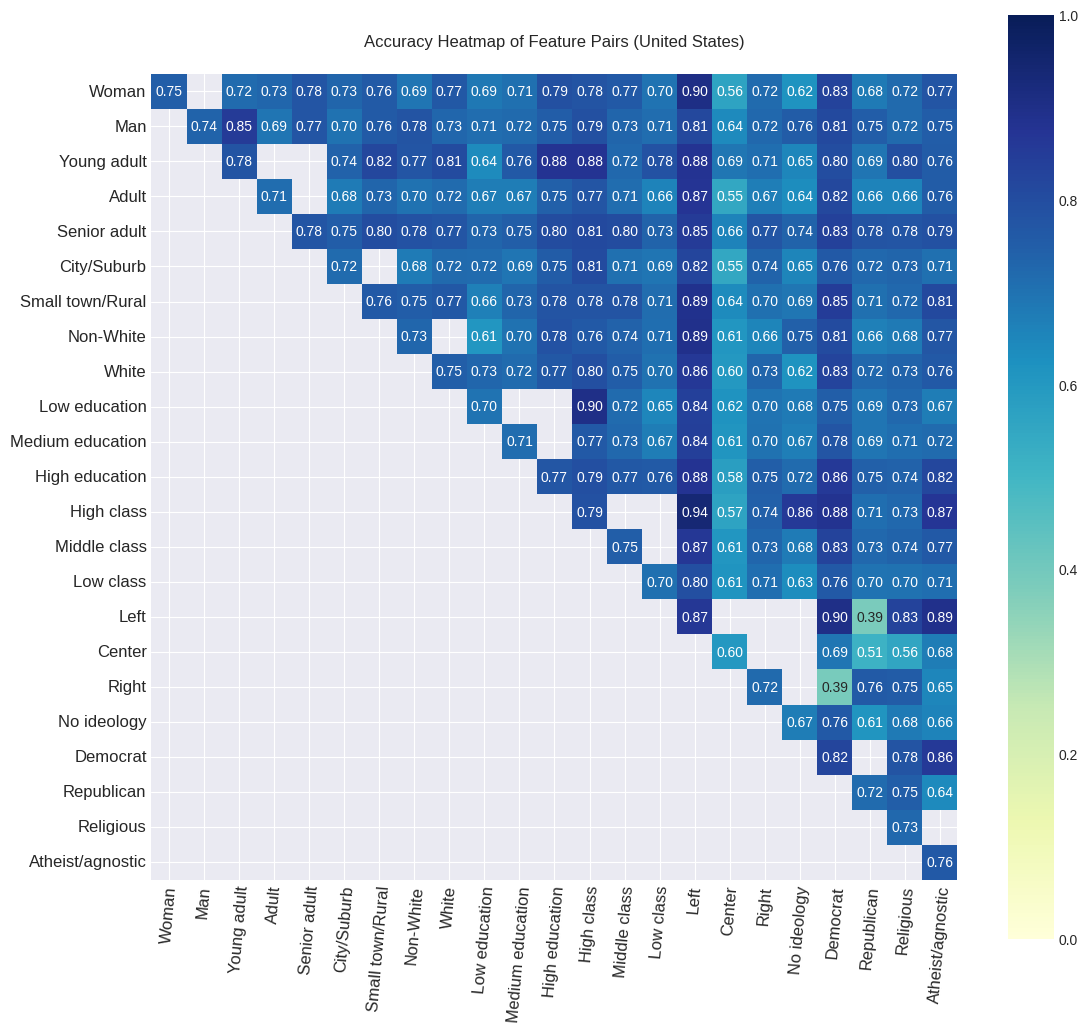} 
     \caption{ Mean accuracy by pairs of socio-demographic groups in the U.S. The matrix shows for each pair of socio-demographic groups the average accuracy obtained between the 3 experiments using the same model and \textit{prompts}.} \label{matriz_acc_eeuu_eng} 
\end{figure}

\subsection{Ablation Analysis of model performance}
\label{sec:ablation_table_appendix}

\begin{table}[h]
\centering
\begin{tabular}{|>{\columncolor[HTML]{EFEFEF}}c|c|c|c|c|c|c|c|c|c|c|}
\hline
\cellcolor[HTML]{C0C0C0}{\color[HTML]{000000}}& \multicolumn{4}{|c|}{\cellcolor[HTML]{C0C0C0}{\color[HTML]{000000} Presidential Voting}} & \multicolumn{4}{|c|}{\cellcolor[HTML]{C0C0C0}{\color[HTML]{000000} Abortion}} & \multicolumn{2}{|c|}{\cellcolor[HTML]{C0C0C0}{\color[HTML]{000000} Constitution }} \\ \cline{2-11} 
\cellcolor[HTML]{C0C0C0}{\color[HTML]{000000}Model }& \multicolumn{2}{|c|}{\cellcolor[HTML]{C0C0C0}{\color[HTML]{000000} U.S.}} & \multicolumn{2}{|c|}{\cellcolor[HTML]{C0C0C0}{\color[HTML]{000000} Chile}} & \multicolumn{2}{|c|}{\cellcolor[HTML]{C0C0C0}{\color[HTML]{000000} U.S.}} & \multicolumn{2}{|c|}{\cellcolor[HTML]{C0C0C0}{\color[HTML]{000000} Chile}} & \multicolumn{2}{|c|}{\cellcolor[HTML]{C0C0C0}{\color[HTML]{000000} Referendum}} \\ \cline{2-11} 
\cellcolor[HTML]{C0C0C0}{\color[HTML]{000000} }     & \cellcolor[HTML]{C0C0C0}{\color[HTML]{000000} Acc} & \cellcolor[HTML]{C0C0C0}{\color[HTML]{000000} JSS} & \cellcolor[HTML]{C0C0C0}{\color[HTML]{000000} Acc} & \cellcolor[HTML]{C0C0C0}{\color[HTML]{000000} JSS} & \cellcolor[HTML]{C0C0C0}{\color[HTML]{000000} Acc} & \cellcolor[HTML]{C0C0C0}{\color[HTML]{000000} JSS} & \cellcolor[HTML]{C0C0C0}{\color[HTML]{000000} Acc} & \cellcolor[HTML]{C0C0C0}{\color[HTML]{000000} JSS} & \cellcolor[HTML]{C0C0C0}{\color[HTML]{000000} Acc} & \cellcolor[HTML]{C0C0C0}{\color[HTML]{000000} JSS} \\ \hline
\textit{All} & 0.88 & 0.92 & 0.69 & 0.87 & 0.63 & 0.91 & 0.51 & 0.78 & 0.63 & 0.91\\ \hline
Without political variables & \textbf{0.61} & \textbf{0.80} & 0.54 & 0.84 & \textbf{0.50} & \textbf{0.72} & 0.47 & 0.78 & 0.49 & 0.85 \\ \hline
Only political variables & 0.87 & \textbf{0.80} & 0.59 & \textbf{0.79} & 0.55 & 0.80 & \textbf{0.43} & \textbf{0.70} & 0.53 & \textbf{0.71} \\ \hline
Without gender & 0.88 & 0.93 & 0.66 & 0.87 & 0.61 & 0.90 & 0.50 & 0.74 & 0.60 & 0.84 \\ \hline
Without Age & 0.88 & 0.92 & 0.67 & 0.85 & 0.62 & 0.90 & 0.47 & 0.73 & 0.58 & 0.83 \\ \hline
Without region & 0.87 & 0.92 & 0.65 & 0.87 & 0.60 & 0.90 & 0.49 & 0.72 & 0.58 & 0.80 \\ \hline
Without race & 0.87 & 0.93 & 0.67 & 0.89 & 0.63 & 0.89 & 0.51 & 0.75 & 0.58 & 0.81 \\ \hline
Without gse & 0.86 & 0.91 & 0.65 & 0.90 &  0.61 & 0.90 & 0.49 & 0.74 & 0.60 & 0.82 \\ \hline
Without scholarity & 0.87 & \textbf{0.80} & 0.66 & 0.81 & 0.60 & 0.90 & 0.50 & 0.75 & 0.59 & 0.81 \\ \hline
Without religion & 0.86 & 0.91 & 0.63 & 0.87 & 0.62 & 0.89 & 0.48 & 0.75 & 0.58 & 0.79 \\ \hline
Without ideology & 0.83 & 0.88 & \textbf{0.53} & 0.80 & 0.59 & 0.91 & 0.47 & 0.74 & \textbf{0.47} & 0.75 \\ \hline
Without party & 0.85 & 0.90 & 0.67 & 0.86 & 0.61 & 0.88 & 0.49 & 0.74 & 0.58 & 0.82 \\ \hline
Without interest & 0.87 & 0.93 & 0.67 & 0.90 & 0.59 & 0.89 & 0.48 & 0.74 & 0.60 & 0.83 \\ \hline
\end{tabular}
\caption{Ablation performance of the Llama-13B model varying the socio-demographic features used for prediction. Each entry contains the mean accuracy (Acc) and JSS values. The lowest values of JSS and accuracy for each experiment are highlighted in bold.}
\label{table:ablation_table_appendix}
\end{table}


The ablation experiments provide valuable insights to clarify whether differences in results are due to the model or the data. Table~\ref{table:ablation_table_appendix} shows performance results varying the socio-demographic features used for prediction. In Chile, using political variables only is insufficient for accurate predictions, achieving less than 60\% precision for the presidential election, plebiscite, and abortion scenarios. In contrast, in U.S. presidential elections, political variables significantly enhance predictions, reaching 87\% precision. On the other hand, in the abortion experiment, a 12\% difference in accuracy is obtained between both countries, again indicating that political variables in the United States are a greater contribution to the model's performance.


When using prompts without political variables, the difference between the two countries in both the presidential and abortion experiments is considerably reduced. This is evidenced by similar Jaccard Similarity Scores (JSS) and a maximum accuracy difference of 6\%. Also, in the U.S. abortion experiment, removing political variables from the prompt results in a greater performance drop compared to Chile, where the variation is nearly negligible. This suggests that LLMs depend on political variables to achieve strong predictive performance. This is not the case in Chile, which has a less politicized society.

\subsection{Prompt Analysis} 
\label{sec:prompting_appendix}
\begin{figure}[t] 
    \centering 
    \includegraphics[scale=0.5]{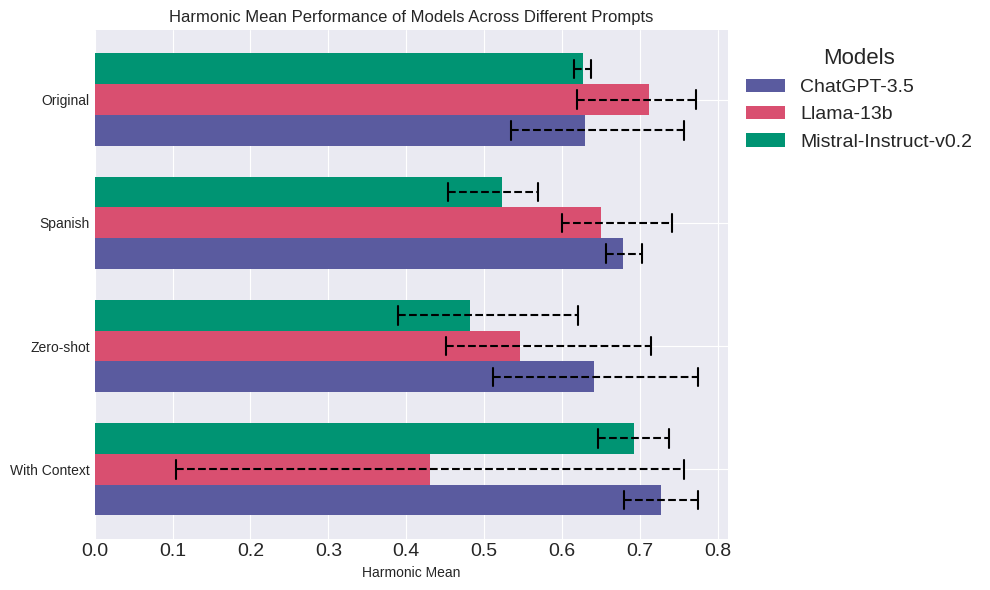}
    \caption{Prompt Sensitivity Analysis Across Models. To construct this graph, the harmonic mean of the results obtained across the different experiments was calculated for each prompt variation. This procedure allows us to visualize the overall performance of the different prompt variations. However, performance differences are not limited to the prompt level; variations are also observed between experiments. That is, a prompt with superior performance for a specific model in a given experiment might exhibit a decrease in performance in another. To represent this variability, intervals are shown, encompassing the minimum and maximum values obtained by each prompt and model variation across the experiments.}
    \label{fig:prompting_appendix}
\end{figure}

The analysis of prompt variations in Figure~\ref{fig:prompting_appendix} reveals that models exhibit differing sensitivities to prompt language, structure, and context. For some models, using Spanish or omitting examples improves performance, while for others, it has the opposite effect. A prompt configuration that excels in one experiment may underperform in another. Interestingly, models demonstrate lower sensitivity to prompts in the U.S. context, suggesting a better understanding of this region. 

Despite these variations, the most consistently effective prompt configuration combines the \textit{chain-of-thought} (CoT) technique, English, and few-shot examples, particularly when applied to the Llama model. This combination appears to provide the optimal balance of clarity and guidance, enhancing prediction accuracy across tasks.

To explore these differences, the experiments employ four distinct prompt variations:

\begin{itemize}
    \item \textbf{Original Prompt:} A baseline prompt with a CoT structure and a few-shot approach using 5 random examples, without any event-specific context.
    \item \textbf{Spanish Prompt:} A translation of the original prompt into Spanish, preserving the CoT structure and few-shot examples, to evaluate performance in a Spanish-speaking context.
    \item \textbf{Zero-Shot Prompt:} A simplified prompt without few-shot examples, testing the model's ability to perform tasks with minimal guidance.
    \item \textbf{Prompt with Context:} Enhances the original prompt by adding event-specific background information, assessing whether additional context improves prediction accuracy.
\end{itemize}

\end{document}